\documentclass[twocolumn,preprintnumbers,floats,prd,amssymb,floatfix,nofootinbib,balancelastpage,superscriptaddress,amsmath]{revtex4-1}
\usepackage[utf8]{inputenc}
\usepackage{CJKutf8}
\usepackage{amssymb}
\usepackage{amsmath}
\usepackage{amsfonts}
\usepackage{graphicx}
\usepackage{color}
\usepackage{xspace}
\usepackage{comment}
\usepackage{hyperref}
\usepackage[normalem]{ulem}
\usepackage[section]{placeins}
\usepackage{afterpage}
\usepackage{slashed}
\usepackage{enumerate}
\usepackage{enumitem}
\usepackage{caption}
\usepackage{subfigure}
\usepackage{float}
\usepackage{ulem}
\usepackage{verbatim}
\usepackage{multirow,rotating}
\usepackage[dvipsnames]{xcolor}
\usepackage{braket}
\usepackage{booktabs} % 提供更好的表格横线
\usepackage{array}     % 提供更灵活的列格式控制

\definecolor{dcolour}{rgb}{.5, .5, .5}
\def\gsim{\raise0.3ex\hbox{$\;>$\kern-0.75em\raise-1.1ex\hbox{$\sim\;$}}}
\def\lsim{\raise0.3ex\hbox{$\;<$\kern-0.75em\raise-1.1ex\hbox{$\sim\;$}}}
\def\gsim{\raise0.3ex\hbox{$\;>$\kern-0.75em\raise-1.1ex\hbox{$\sim\;$}}}
\def\lsim{\raise0.3ex\hbox{$\;<$\kern-0.75em\raise-1.1ex\hbox{$\sim\;$}}}

\newcommand{\ba}[1]{\begin{eqnarray} \label{(#1)}}
	\newcommand{\ea}{\end{eqnarray}}

\newcommand{\Eq}[1]{Eq.~\ref{#1}}
\newcommand{\FIG}[1]{FIG.~\ref{#1}}
\newcommand{\TAB}[1]{Table~\ref{#1}}
\begin{document}
\captionsetup[figure]{justification=raggedright,singlelinecheck=false}
\title{Theoretical study of $f_0(980)$, $a_0(980)$ and $\Xi(1/2^-)$ in the process $\Xi_c^+ \to \Sigma^+K^+K^-$}
\author{Ruitian Li}
\email{liruitian@mail.dlut.edu.cn}
\affiliation{Institute of Theoretical Physics, School of Physics, Dalian University of Technology, \\ 
		No.2 Linggong Road, Dalian, Liaoning, 116024, People’s Republic of China}
\author{Xuan Luo}
\email{xuanluo@ahu.edu.cn}
\affiliation{School of Physics and Optoelectronics Engineering, Anhui University, \\
		Hefei, Anhui 230601, People’s Republic of China}
\author{Hao Sun}
\email{haosun@dlut.edu.cn}
\affiliation{Institute of Theoretical Physics, School of Physics, Dalian University of Technology, \\ 
	No.2 Linggong Road, Dalian, Liaoning, 116024, People’s Republic of China}
\begin{abstract}
	We employed the chiral unitarity approach to investigate the decay process $\Xi_c^+ \to \Sigma^+K^+K^-$, by considering that the  low lying nucleon resonance $\Xi(1/2^-)$ and the  low lying scalar meson $f_0(980)$ and $a_0(980)$ that could be dynamically generated through $S$-wave pseudoscalar meson-octet baryon and the $S$-wave pseudoscalar meson-pseudoscalar meson interactions, respectively. In the invariant mass distributions of $\Sigma^+K^-$ and $K^+K^-$, we observe a distinct peak structure associated with the resonant state $\Xi(1/2^-)$ and a bit enhancement near the $K^+K^-$ threshold that is corresponding to the mesons $f_0(980)$ and $a_0(980)$, respectively. Consequently, we recommend more precise experimental measurements of this process in the future.
\end{abstract}
%%%%%%%%%%%%%%%%%%%%%%%%%%%%%%%%%%%%%%%%%%%%%%%%%%%%%%%%%%%%%%%%%%%%%%
%\arxivnumber{}
%\pacs{}
\vskip10 mm
\maketitle
\flushbottom	

\section{Introduction}
\label{I}

The discovery of low-lying excited baryons has attracted broad attention. However, further research is still needed to clarify  their internal structure. Research on the properties of $\Xi$ states has attracted much attention in hadron physics~\cite{Garcia-Recio:2003ejq,Gamermann:2011mq,Pervin:2007wa,PavonValderrama:2011gp,Nishibuchi:2022zfo}.

Apart from the ground states of the $\Xi(1321)$ and $\Xi(1530)$ with spin-parity $J^P=1/2^+$ and $J^P=3/2^+$, respectively, which are well established with four-star ratings, the situation of other $\Xi$ excited states is still unclear, with ratings of less than three stars~\cite{ParticleDataGroup:2024cfk}.

In recent years, many developments have been made in obtaining detailed experimental data on low-lying $\Xi$ excited states. 
In 2002, the Belle Collaborations have first reported the evidence of the resonant contribution $\Lambda_c^+ \to \Xi(1690)^0K^+$~\cite{Belle:2001hyr}, then later confirmed by the FOCUS and BaBar
Collaboration~\cite{FOCUS:2005sye,BaBar:2006tck}.
The Belle collaboration, in 2019, reported distinct peaks in the $\pi^+\Xi^-$ invariant mass distribution from the $\Xi_c^+ \to \Xi^-\pi^+\pi^+$ decay, which are correlated with the $\Xi(1620)$ and $\Xi(1690)$ states~\cite{Belle:2018lws}.
Then in 2020, the $\Xi(1690)$ state was subsequently observed by the LHCb collaboration in the $\Lambda K^-$ invariant mass spectrum of the decay $\Xi_b \to J/\psi \Lambda K^-$~\cite{LHCb:2020jpq}.
Recently, the BESIII collaboration has discovered signals of $\Xi(1690)$ and $\Xi(1820)$ in the $\psi(3686) \to K^-\Lambda \bar{\Xi}^+$ reaction, determining their spin as $J^P =1/2^-$ and $3/2^-$, respectively~\cite{BESIII:2023mlv}. 

In theoretical research, the nature of the $\Xi(1/2^-)$ state remains a subject of ongoing debate. 
The Skyrme model~\cite{Oh:2007cr} suggests assigning the $\Xi(1620)$ to a $J^P=1/2^-$ state and also predicts two $\Xi(1/2^-)$ states with masses of $1616$ MeV and $1658$ MeV. Many studies have explored $\Xi$ states using the quark model~\cite{Capstick:1986ter,Glozman:1995fu,Melde:2008yr,Yan:2024usf}, their properties have also been studied via lattice QCD~\cite{Engel:2013ig}, QCD sum rules~\cite{Jido:1996zw,Lee:2002jb,Aliev:2018hre}, and large-$N_c$ analysis~\cite{Schat:2001xr,Goity:2003ab,Matagne:2004pm,Matagne:2006zf,Semay:2007cv}.
There are also many studies~\cite{Ramos:2002xh,Miyahara:2016yyh,Garcia-Recio:2003ejq,Li:2025exm} that, through the chiral unitary approaches, generated the $J^P=1/2^-$ states $\Xi(1620)$ and $\Xi(1690)$ via the interaction of the coupled channels $\pi\Sigma$, $\bar{K}\Lambda$, $\bar{K}\Sigma$, and $\eta\Xi$. Recent Refs.~\cite{Liu:2023jwo,Magas:2024mba} investigating the $\Lambda_c^+\to \Lambda K^+\bar{K}^0$ and $\Xi_c^+\to \Xi^-\pi^+\pi^-$ processes measured by the Belle experiment support the molecular interpretation of the $\Xi(1690)$ state.

Moreover, light scalar mesons are difficult to identify due to their large decay widths, and their structure remains subject to various interpretations, such as the traditional $q\bar{q}$ state,  hadronic molecule or multi-quark states, etc~\cite{Klempt:2007cp,Close:2002zu,Amsler:2004ps,Bugg:2004xu,Pelaez:2015qba}. While the $a_0(980)$ and $f_0(980)$ states have been discovered many years, their internal structures remain an active topic of research~\cite{Astier:1967zz,Ammar:1968zur,Defoix:1968hip,ParticleDataGroup:2024cfk,Oller:1997ng,Albuquerque:2023bex,Baru:2003qq}. In experiments, the resonance states $a_0(980)$ and $f_0(980)$ cannot be distinguished because they share the same quantum number $J^{PC}=0^{++}$ and exhibit strong overlap in their decay to the $K^+K^-$ channel.

%Despite $a_0(980)$ with isospin $I = 1$ and $f_0(980)$ with $I = 0$ both masses around 980 MeV and slightly below the $K\bar{K}$ threshold, they cannot be experimentally distinguished. This is because their quantum number $J^{PC} = 0^{++}$ are same and exhibit strong overlap in decaying to $K^+K^-$.

The isospin of $a_0(980)$ is $I=1$, while that of $f_0(980)$ is $I=0$. Both have masses of about 980 MeV, slightly below the $K\bar{K}$ threshold. Many studies have explored the charm hadrons decay processes, in which $a_0(980)$ and $f_0(980)$ are dynamically generated through final-state interactions within the chiral unitary approach~\cite{Liang:2016hmr,Debastiani:2016ayp,Oset:2016lyh,Wang:2020pem,Feng:2020jvp,Rahmani:2025uut,Song:2025dgg,Li:2025msk}.

Therefore, in this work, we will analyze the process $\Xi_c^+ \to \Sigma^+K^+K^-$ using the chiral unitary approach considering the $S$-wave pseudosalar meson-octet baryon interaction, which will dynamically generate the $\Xi(1/2^-)$ states. Moreover, we also taking into account the contribution of scalar meson $a_0(980)$ and $f_0(980)$
from the $S$-wave pseudosalar meson-pseudosalar meson interaction. Then, we calculate the $K^-\Sigma^+$ and $K^+K^-$ invariant mass distributions in the $\Xi_c^+ \to \Sigma^+K^+K^-$ reaction. The work done here
should be an incentive for this more accurate experimental
analysis to be performed.

This paper is organized as follows. First, in Sec.~\ref{II}, we present the theoretical formalism of the process $\Xi_c^+ \to \Sigma^+K^+K^-$. Then, in Sec.~\ref{III}, we discuss numerical results and discussions. Finally, we provide a brief summary in Sec.~\ref{IV}.

\section{formalism}
\label{II}

In this section, we present the theoretical formalism for the process $\Xi_c^+ \to \Sigma^+K^+K^-$.  First, the mechanism for the process via the intermediate state $\Xi(1/2^-)$ is presented in Subsec.~\ref{II.a}. Subsequently, in Subsec.~\ref{II.b}, we present the $S$-wave final state interactions of $K^+K^-$, wich dynamically generate the scalar resonances $f_0(980)$ and $a_0(980)$.
\subsection{The  $\Xi(1/2^-)$ role in $\Xi_c^+ \to \Sigma^+K^+K^-$}
\label{II.a}

\begin{figure}[tbhp]
	\centering
	\includegraphics[scale=0.6]{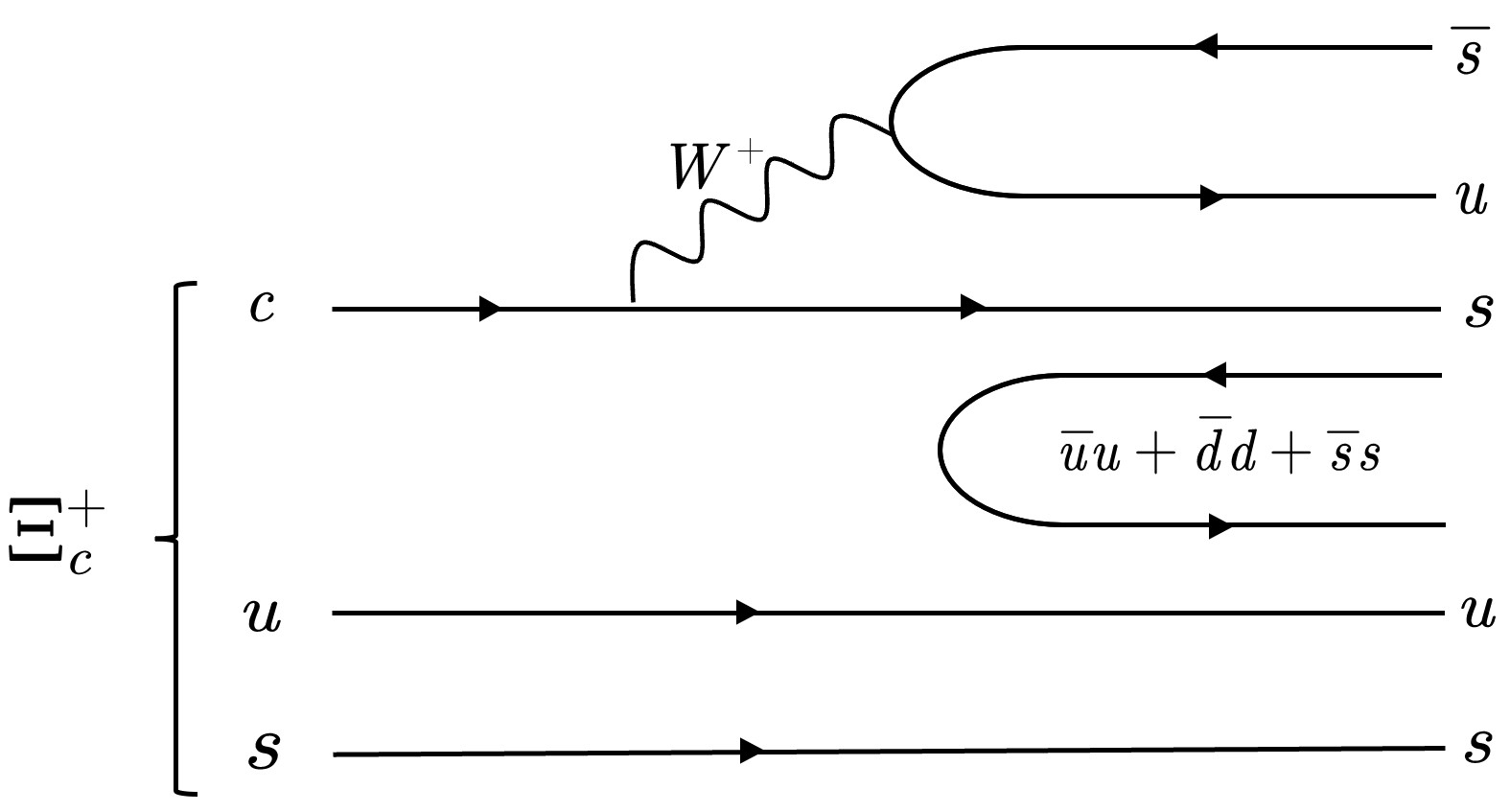}
	\caption{Quark level diagram for the process $\Xi_c^+ \to K^+ s \left(\bar{u}u+\bar{d}d+\bar{s}s\right)us$ via the $W^+$ external emission.}
	\label{fig1a}
\end{figure}

\begin{figure}[tbhp]
	\centering
	\includegraphics[scale=0.6]{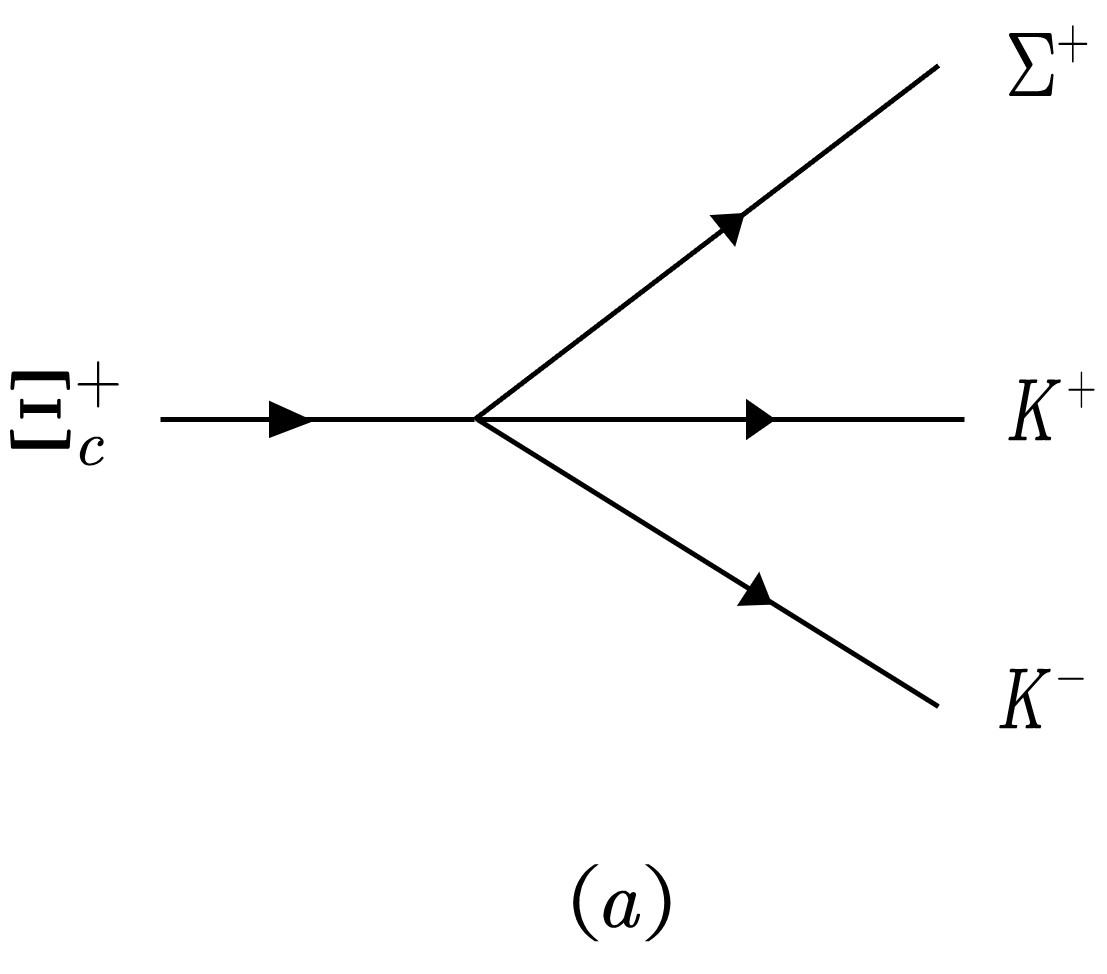}\\
	\includegraphics[scale=0.6]{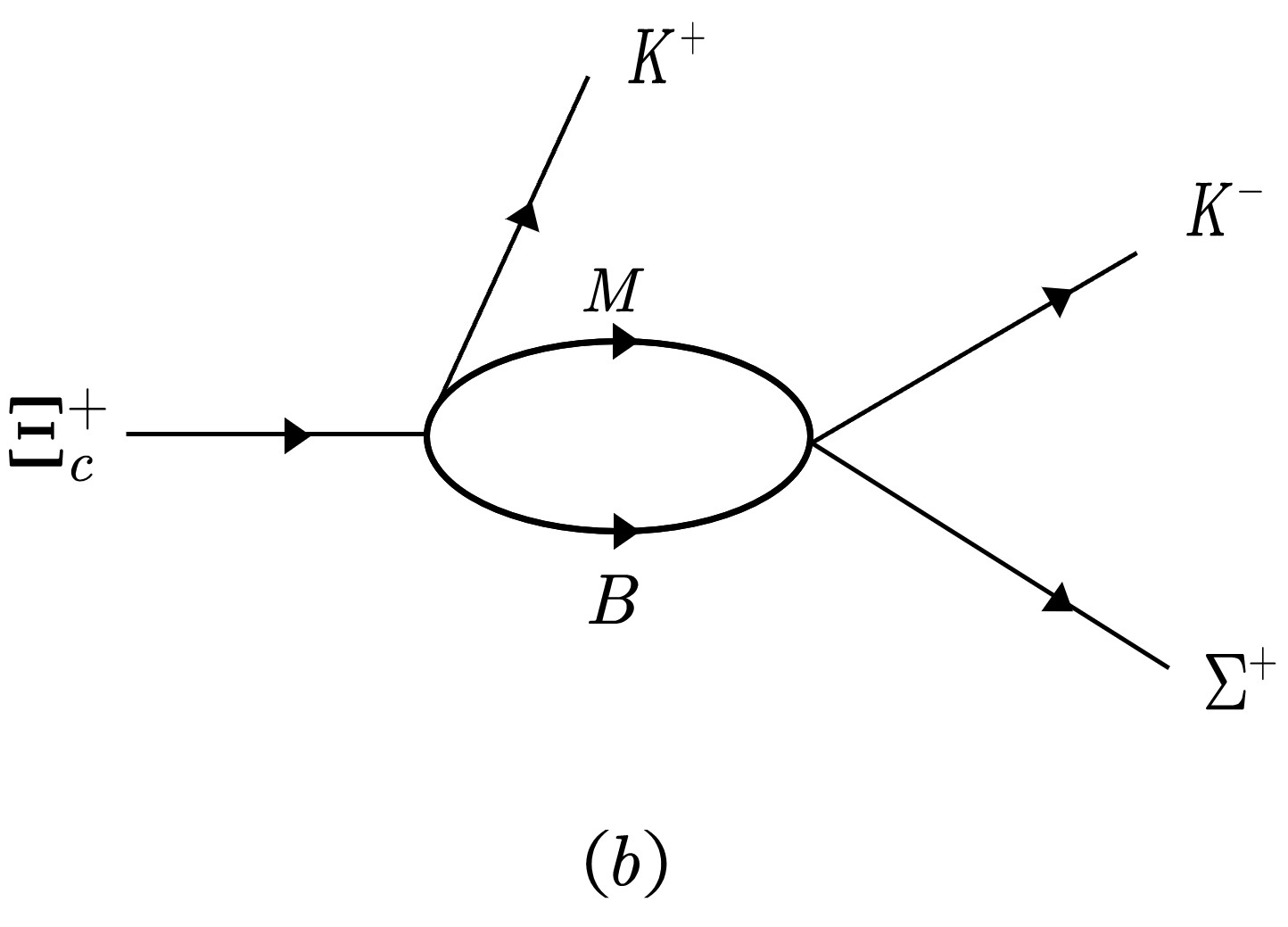}
	\caption{The mechanisms of the decay $\Xi_c^+ \to \Sigma^+K^+K^-$, (a) tree diagram, (b) the $S$-wave final state interactions.}
	\label{fig1b}
\end{figure}

In this work, for the $\Xi_c^+ \to K^+(MB)\to K^+K^-\Sigma^+$ process, we consider the $W^+$ external emission as shown in \FIG{fig1a}.
For \FIG{fig1a}, the initial state $\Xi_c^+$ of the $c$ quark first weakly decays into a $s$ quark and a $W^+$ boson. Subsequently, $W^+$ boson weakly decays to the $u\bar{s}$ quark pair.  The $u\bar{s}$ quark pair is then hadronized to $K^+$, while the remaining quark cluster $sus$ and the quark pair $\bar{q}q=\bar{u}u+\bar{d}d+\bar{s}s$ produced in the vacuum with quantum numbers $J^{PC}=0^{++}$ hadronize into meson-baryon(MB) pairs. This processe can be expressed as follows:
\begin{equation}
\begin{aligned}
\Xi_c^+&\Rightarrow \frac{1}{\sqrt{2}}c\left(su-us\right)\\
&\Rightarrow V_{cs}\frac{1}{\sqrt{2}}W^+ s\left(su-us\right)\\
& \Rightarrow V_{cs}V_{us} \frac{1}{\sqrt{2}}u\bar{s}s\left(u\bar{u}+d\bar{d}+s\bar{s}\right)\left(su-us\right)\\
& \Rightarrow V_{cs}V_{us}\frac{1}{\sqrt{2}}K^+ s\left(u\bar{u}+d\bar{d}+s\bar{s}\right)\left(su-us\right)
\end{aligned}
\end{equation}
where we use the baryon flavor wave function $\Xi_c^+ = \frac{1}{\sqrt{2}}c\left(su-us\right)$, and the CKM matrix elements related to the
Cabibbo angle. We take $V_{cd}=V_{us} =-\sin\theta_c=-0.22534$ and $V_{ud}=V_{cs}=\sin\theta_c=0.97427$~\cite{Wang:2020pem,Feng:2020jvp}. Based on the $SU(3)$ flavor symmetry, we connect the two degrees of freedom of quarks and hadrons using the meson matrix $M$ and the baryon matrix $B$. The forms of the $M$ and $B$ matrices~\cite{Bramon:1992kr,Molina:2019udw,Lyu:2023aqn} are as follows:
\begin{eqnarray}
	M&=&\left(\begin{matrix} u\bar{u} & u\bar{d} & u\bar{s}  \\
		d\bar{u}  &   d\bar{d}  &  d\bar{s} \\
		s\bar{u}  &  s\bar{d}   &    s\bar{s}
	\end{matrix}
	\right)   \\  
	&=&\left(\begin{matrix} \frac{\eta}{\sqrt{3}}+ \frac{{\pi}^0}{\sqrt{2}}+ \frac{{\eta}'}{\sqrt{6}} & \pi^+ & K^+  \\
		\pi^-  &   \frac{\eta}{\sqrt{3}}- \frac{{\pi}^0}{\sqrt{2}}+ \frac{{\eta}'}{\sqrt{6}}  &  K^0 \\
		K^-  &  \bar{K}^{0}   &    -\frac{\eta}{\sqrt{3}}+ \frac{{\sqrt{6}\eta}'}{3}
	\end{matrix}
	\right),\nonumber
\end{eqnarray}
\begin{eqnarray}
B &=& \frac{1}{\sqrt{2}}\begin{pmatrix} 
u(ds-sd) & u(su-us) & u(ud-du)  \\
d(ds-sd) & d(su-us) & d(ud-du) \\
s(ds-sd) & s(su-us) & s(ud-du)
\end{pmatrix} \\
&=& \begin{pmatrix} 
\frac{\Sigma^0}{\sqrt{2}} + \frac{{\Lambda}}{\sqrt{6}} & \Sigma^+ & p  \\
\Sigma^- & -\frac{\Sigma^0}{\sqrt{2}} + \frac{{\Lambda}}{\sqrt{6}}  & n \\
\Xi^-  & \Xi^{0} & -\frac{2\Lambda}{\sqrt{6}} 		\end{pmatrix},
\nonumber
\end{eqnarray}
where we have employed the $\eta-\eta'$ mixing as in Refs.~\cite{Bramon:1994cb,Lyu:2024qgc} and the $\eta-\eta'$ mixing angle $\sin\theta = -\frac{1}{3}$, which is a standard choice~\cite{Bramon:1994cb}. The hadronization processes can be expressed using the meson matrix $M$ and baryon matrix $B$:
\begin{equation}
\begin{aligned}
H_1 &\Rightarrow V_{cs}V_{us}\frac{1}{\sqrt{2}}K^+ s\left(u\bar{u}+d\bar{d}+s\bar{s}\right)\left(su-us\right) \\
 &= V_{cs}V_{us}K^+ \sum M_{3i}B_{i2}\\
 &=V_{cs}V_{us} K^+ \left( |K^-\Sigma^+\rangle -\frac{1}{\sqrt{2}}|\bar{K}^0\Sigma^0\rangle +\frac{1}{\sqrt{6}}|\bar{K}^0 \Lambda \rangle \right.\\ 
 &\left. -\frac{1}{\sqrt{3}}|\eta\Xi^0\rangle\right).
\end{aligned}
\end{equation}

Thus, the process $\Xi_c^+ \to \Sigma^+K^+K^-$ could occure via the tree diagram of \FIG{fig1b}(a) and the $S$-wave pseudoscalar meson-octet baryon interactions of \FIG{fig1b}(b). The corresponding decay amplitude in \FIG{fig1b} is:
\begin{equation}
\begin{aligned}
\mathcal{T}^{\Xi(1/2^-)}_1 & = C V_p V_{cs}V_{us} \\
&\times \left[ 1+ G_{K^-\Sigma^+}(M_{K^-\Sigma^+}) t_{K^-\Sigma^+ \to K^-\Sigma^+}(M_{K^-\Sigma^+}) \right.\\ 
&\left. -\frac{1}{\sqrt{2}}G_{\bar{K}^0\Sigma^0}(M_{K^-\Sigma^+})t_{\bar{K}^0\Sigma^0 \to K^-\Sigma^+ }(M_{K^-\Sigma^+}) \right.\\
&\left. +\frac{1}{\sqrt{6}}G_{\bar{K}^0\Lambda}(M_{K^-\Sigma^+})t_{\bar{K}^0\Lambda \to K^-\Sigma^+} (M_{K^-\Sigma^+})\right. \\
&\left. -\frac{1}{\sqrt{3}} G_{\eta \Xi^0}(M_{K^-\Sigma^+}) t_{\eta\Xi^0 \to K^-\Sigma^+}(M_{K^-\Sigma^+})
\right],
\end{aligned}\label{eqt1}
\end{equation}
where the color factor $C$ represents the relative weight of the external $W^+$ emission mechanism in \FIG{fig1a} compared to internal $W^+$ emission, the parameter $V_p$ represents the strength of the weak interaction vertex in \FIG{fig1a}. The focus of this work is on the final state interactions of this process, so we assume $V_p$ to be constant.
 $G$ is the meson-baryon loop function. Typically, we may adopt either the three-momentum cutoff method or dimensional regularization scheme~\cite{Inoue:2001ip,Guo:2005wp}. Here, we take the dimensional regularization method to obtain the following analytic form for loop function $G$,
\begin{equation}
\begin{aligned}
G_{MB} &= \frac{2M_B}{16\pi^2} \Bigg\{ a_{MB}(\mu) + \ln\frac{M_B^2}{\mu^2} + \frac{m_M^2 - M_B^2 + s}{2s} \ln\frac{m_M^2}{M_B^2} \\
& + \frac{q_{MB}}{\sqrt{s}} \bigg[ \ln\left(s - M_B^2 + m_M^2 + 2q_{MB}\sqrt{s}\right) \nonumber \\
&\quad + \ln\left(s + M_B^2 - m_M^2 + 2q_{MB}\sqrt{s}\right)  \\
& - \ln\left(-s + M_B^2 - m_M^2 + 2q_{MB}\sqrt{s}\right) \\
& - \ln\left(-s - M_B^2 + m_M^2 + 2q_{MB}\sqrt{s}\right) \bigg] \Bigg\},
\label{eq:G}
\end{aligned}
\end{equation}
where $a_{MB}\left(\mu\right)$ is the subtraction constant, and $\mu$ is a scale of dimensional regularization. $m_M$ and $M_B$ represent the masses of the meson and baryon, respectively.
$q_{MB}$ is the on-shell meson momentum in the center-of-
mass frame of the meson-baryon system, given by 
\begin{equation}
q_{MB} = \frac{\lambda^{1/2}(s,m_M^2, M_B^2)}{2\sqrt{s}}
\end{equation}
with $\lambda(x,y,z) =x^2+y^2+z^2-2xy-2yz-2zx$.

The two-body scattering amplitude $t_{MB-K^-\Sigma^+}$ of the coupled channel in \Eq{eqt1} is calculated by solving the Bethe-Salpeter equation through the method of chiral unitary approach:
\begin{equation}
	T=[1-VG]^{-1}V.
	\label{TT}
\end{equation}
We have included six coupled channels in our calculation: $K^-\Sigma^+$, $\bar{K}^0\Sigma^0$, $\bar{K}^0\Lambda$, $\pi^+\Xi^-$, $\pi^0\Xi^0$ and $\eta\Xi^0$.
The $V$ is transition potential, which is expressed as in Ref.~\cite{Oset:2001cn,Wang:2015pcn},
\begin{equation}
	\begin{aligned}
		V_{ij}=&-C_{ij}\frac{1}{4f_if_j}(2\sqrt{s}-M_i-M_j)\sqrt{\frac{E_i+M_i}{2M_i}}\sqrt{\frac{E_j+M_j}{2M_j}},
	\end{aligned} 
\end{equation}
where, $M_{i(j)}$ and $E_{i(j)}$ represent the mass and energy of the baryon in the $i(j)$-th channel, respectively, and $E_i = (s+M_i^2-m_i^2)/2\sqrt{s}$. The coefficient $C_{ij}$, which reflect $SU(3)$ flavor symmetry, is listed in \TAB{tab1}. The decay constant $f_{i(j)}$ for the meson in the $i(j)$-th channel is:
\begin{equation}
	\begin{aligned}
			f_{\pi}=93\ \text{MeV}, \qquad f_{K} = 1.22f_{\pi}, \qquad f_{\eta} = 1.3f_{\pi} .
	\end{aligned}
\end{equation}
As used in Refs.~\cite{10.1093/ptep/ptv129,Li:2025exm}, We also adopt the parameters for the loop function $G$: $a_{\bar{K}\Sigma} = -1.98$, $a_{\bar{K}\Lambda} = -2.07$, $a_{\pi\Xi} = -0.75$, $a_{\eta\Xi} = -3.31$, and $\mu = 630$ MeV.

\begin{table}[htbp]
	\centering
	\caption{$C_{ij}$ coefficients in the potential for the coupled channels with $S=-2$ $\left(C_{ij} = C_{ji}\right)$~\cite{10.1093/ptep/ptv129}.}
	\setlength{\tabcolsep}{1.5pt}     
	\begin{tabular}{ccccccc}
		\toprule 
		& $K^-\Sigma^+$ & $\bar{K}^0\Sigma^0$ & $\bar{K}^0\Lambda$ & $\pi^+ \Xi^-$& $\pi^0\Xi^0$& $\eta\Xi^0$\\
		\midrule 
		$K^-\Sigma^+$  & 1& $-\sqrt{2}$& 0& 0& $-1/\sqrt{2}$& $-\sqrt{3/2}$  \\
		 $\bar{K}^0\Sigma^0$ & & 0 & 0&$-1/\sqrt{2}$&$-1/2$&$\sqrt{3/4}$    \\
	$\bar{K}^0\Lambda$& & & 0&$-\sqrt{3/2}$&$\sqrt{3/4}$&$-3/2$\\
	$\pi^+ \Xi^-$ & & & &1&$-\sqrt{2}$&0\\
	$\pi^0\Xi^0$ & & & & & 0&0\\
	$\eta\Xi^0$& & & & & & 0\\
		\bottomrule
	\end{tabular}
	\label{tab1}
\end{table}
{\tiny }

\subsection{$S$-wave final state interactions of $K^+K^-$}
\label{II.b}
\begin{figure}[tbhp]
	\centering
	\includegraphics[scale=0.5]{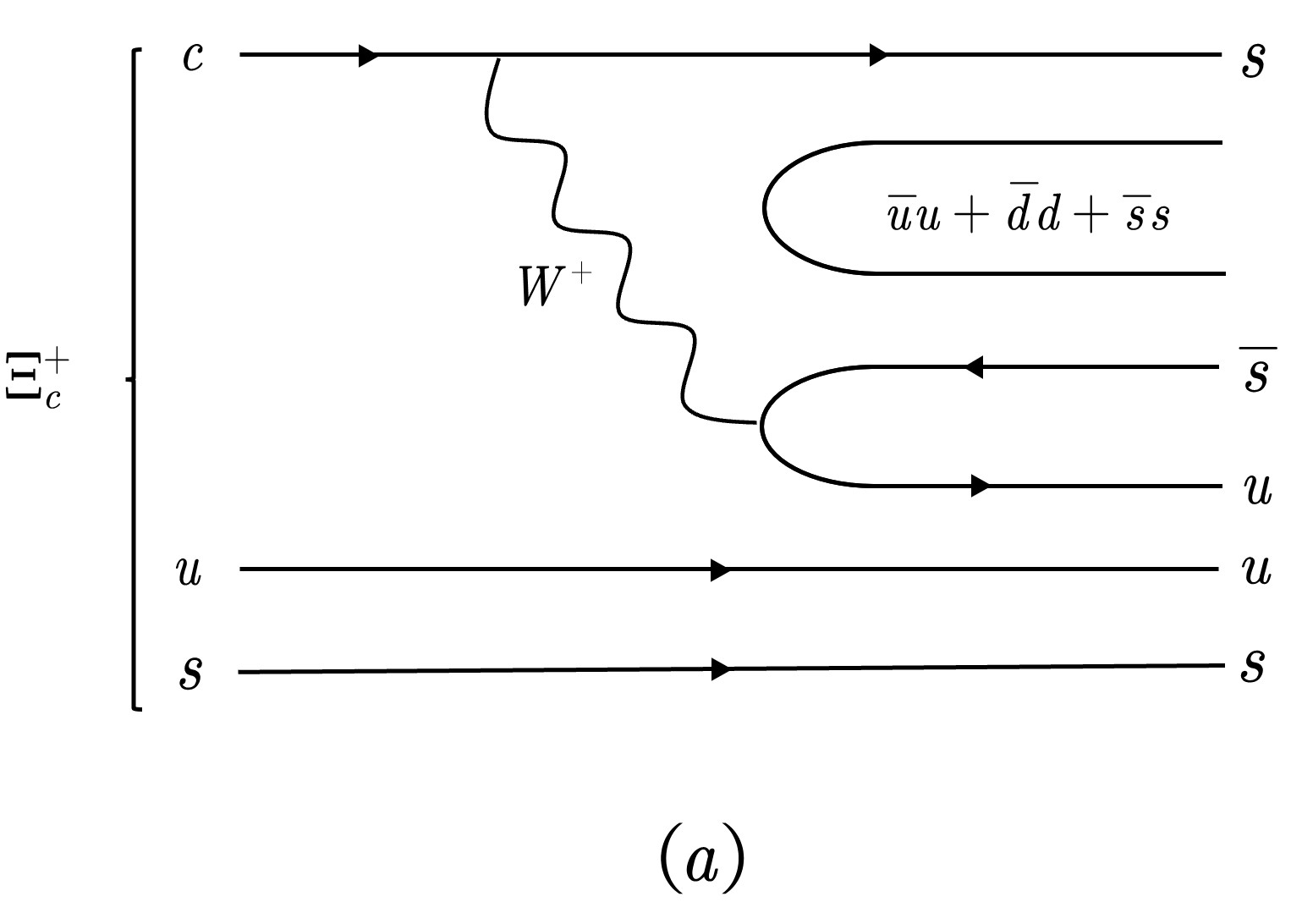}\\
	\includegraphics[scale=0.5]{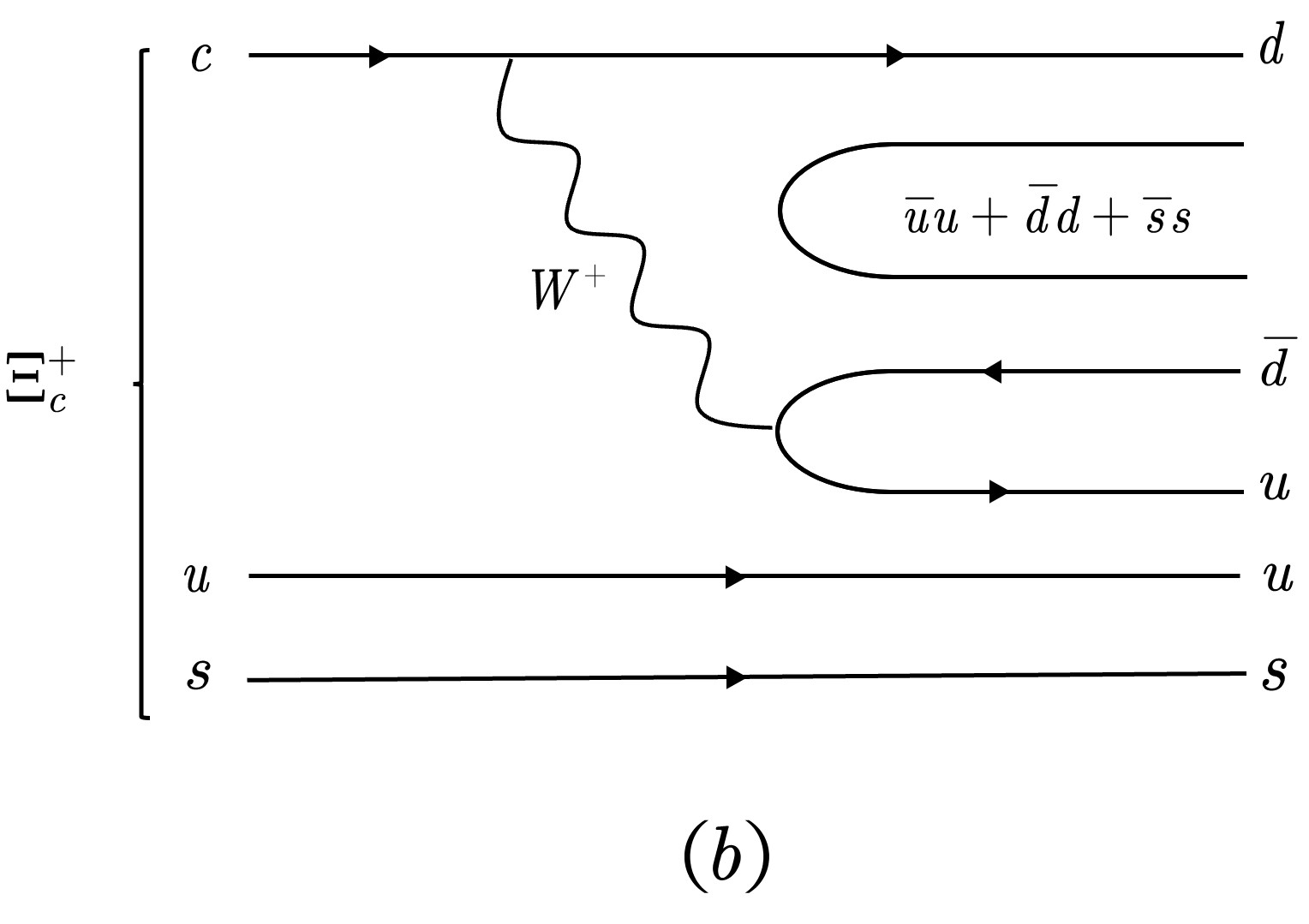}
	\caption{Quark level diagram. (a) The process  $\Xi_c^+\to \Sigma^+ $\\$  s \left(\bar{u}u+\bar{d}d+\bar{s}s\right)\bar{s}$ via the $W^+$ internal emission. (b) Th\\-e process  $\Xi_c^+ \to \Sigma^+ d \left(\bar{u}u+\bar{d}d+\bar{s}s\right)\bar{d}$ via the $W^+$ inter\\-nal emission. }
	\label{fig2a}
\end{figure}

In this subsection, we discuss the theoretical mechanism
of the dynamic generation of  $f_0(980)$ and $a_0(980)$ in the process $\Xi_c^+\to \Sigma^+K^+K^-$, which can happen via the internal $W^+$ emission diagram, as shown in \FIG{fig2a}.

For the $W^+$ internal emission, as shown in
\FIG{fig2a}(a),(b). The $c$ quark in the $\Xi^+_c$ decays into an $s$ (or $d$) quark and a $W^+$ boson via the weak interaction, then the $W^+$ boson decays into an $\bar{s}u$ (or  $\bar{d}u$) pair. The $u$  quark from the $W^+$ decay and the $us$ quark pair of the intial $\Xi_c^+$ will hadronize into a $\Sigma^+$ baryon, while the remaining $s\bar{s}$ (or $d\bar{d}$) quark pair, together with the quark pair $\bar{q}q$ from the vacuum, hadronize into hadron pairs,
\begin{equation}
\begin{aligned}
H_{(2a)}&\Rightarrow \frac{1}{\sqrt{2}}c\left(su-us\right)\\
&\Rightarrow \frac{1}{\sqrt{2}}V_{cs}W^+ s\left(su-us\right)\\
& \Rightarrow \frac{1}{\sqrt{2}}V_{cs}V_{us} s\left( u\bar{u}+d\bar{d}+s\bar{s}\right) \bar{s}u\left(su-us\right)\\
& \Rightarrow V_{cs}V_{us}\Sigma^+ s\left(u\bar{u}+d\bar{d}+s\bar{s}\right)\bar{s}\\
H_{(2b)}&\Rightarrow \frac{1}{\sqrt{2}}c\left(su-us\right)\\
&\Rightarrow \frac{1}{\sqrt{2}}V_{cd}W^+ d\left(su-us\right)\\
& \Rightarrow\frac{1}{\sqrt{2}} V_{cd}V_{ud} d\left( u\bar{u}+d\bar{d}+s\bar{s}\right) \bar{d}u\left(su-us\right)\\
& \Rightarrow V_{cd}V_{ud}\Sigma^+ d\left(u\bar{u}+d\bar{d}+s\bar{s}\right)\bar{d}.
\end{aligned}
\end{equation}
The hadronization processes can be expressed using:
\begin{equation}
\begin{aligned}
H_{(2a)}&\Rightarrow V_{cs}V_{us}\Sigma^+ s\left(u\bar{u}+d\bar{d}+s\bar{s}\right)\bar{s}\\
& \Rightarrow V_{cs}V_{us}\Sigma^+\sum_i \left( M_{3i} \times M_{i3}\right)\\
&  \Rightarrow V_{cs}V_{us}\Sigma^+ \left(|K^+K^- \rangle +|K^0\bar{K}^0\rangle  +\frac{1}{3}|\eta\eta\rangle\right)\\
H_{(2b)}&\Rightarrow    V_{cd}V_{ud}\Sigma^+ d\left(u\bar{u}+d\bar{d}+s\bar{s}\right)\bar{d}\\
& \Rightarrow   V_{cd}V_{ud}\Sigma^+\sum_i \left( M_{2i} \times M_{i2}\right)\\
&  \Rightarrow   V_{cd}V_{ud}\Sigma^+ \left(|\pi^+\pi^-\rangle+ \frac{1}{3} |\eta\eta\rangle  +\frac{1}{2}|\pi^0\pi^0\rangle \right.\\
&\left.-\frac{2}{\sqrt{6}}|\pi^0\eta\rangle +|K^0\bar{K}^0\rangle\right).
\end{aligned}
\end{equation}

\begin{figure}[tbhp]
	\centering
	\includegraphics[scale=0.5]{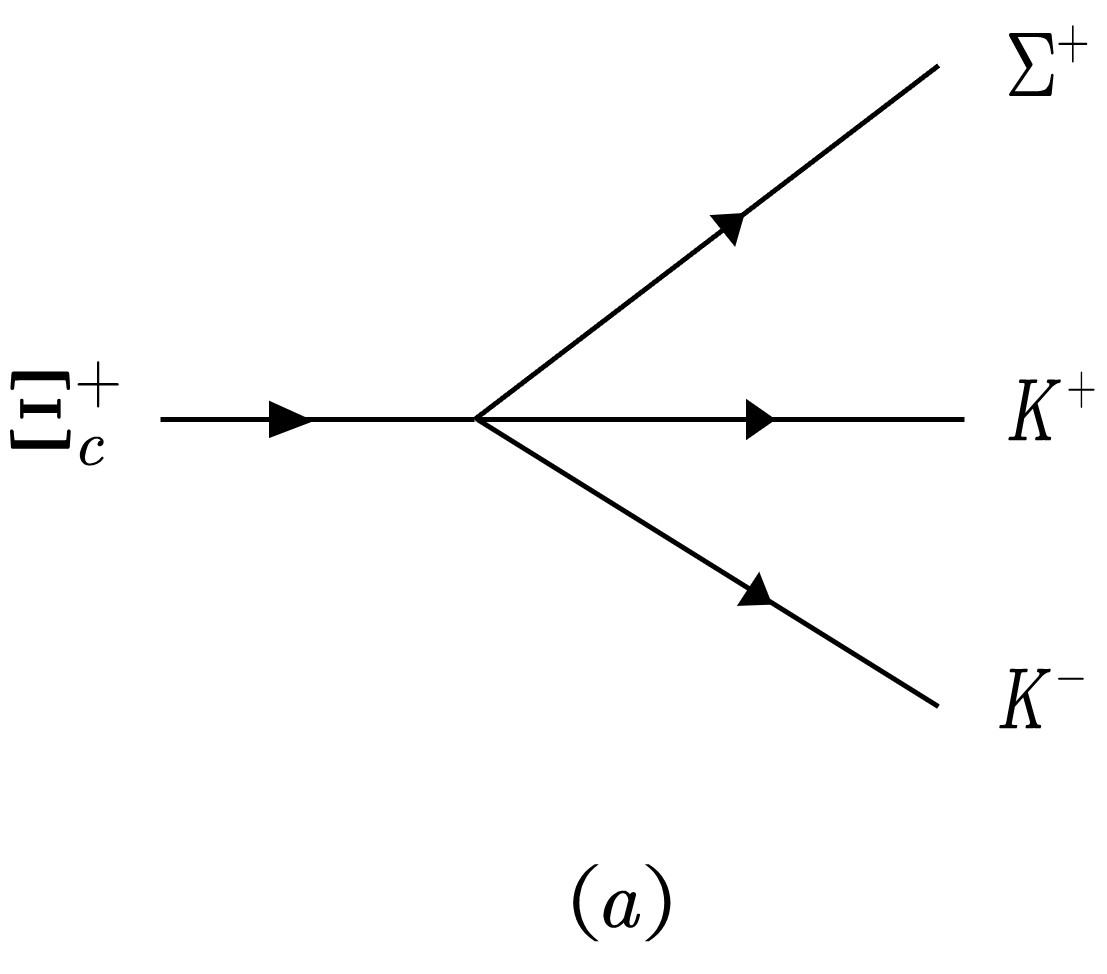}\\
	\includegraphics[scale=0.5]{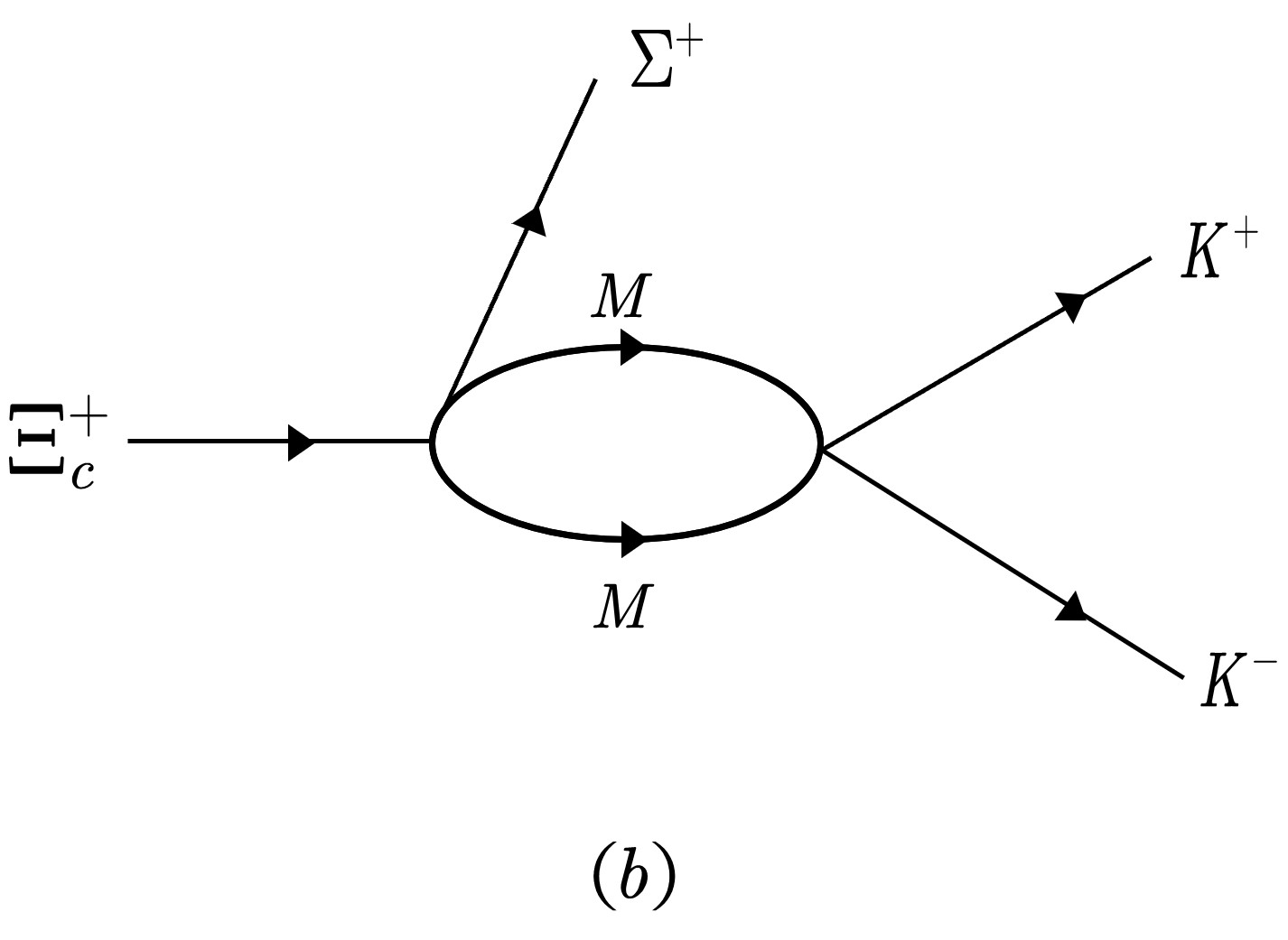}
	\caption{The mechanisms of the decay $\Xi_c^+ \to \Sigma^+K^+K^-$, (a) tree diagram, (b) the $S$-wave final state interactions.}
	\label{fig2b}
\end{figure}

The amplitude contribution of the process $\Xi_c^+ \to \Sigma^+K^+K^-$ for diagram \ref{fig2b}:

\begin{equation}
\begin{aligned}
\mathcal{T}_{2a} =& V_P V_{cs}V_{us} [1+ G_{K^+K^-}(M_{K^+K^-}) t_{K^+K^-\to K^+K^-}(M_{K^+K^-}) \\
& +G_{K^0\bar{K}^0}(M_{K^+K^-}) t_{K^0\bar{K}^0 \to K^+K^-}(M_{K^+K^-})\\
& \left.+\frac{1}{3} G_{\eta\eta}(M_{K^+K^-})\tilde{t}_{\eta\eta \to K^+K^-}(M_{K^+K^-}) \right]\\
\mathcal{T}_{2b} =&V_PV_{cd}V_{ud} [ G_{\pi^+\pi^-}(M_{K^+K^-}) t_{\pi^+\pi^-\to K^+K^-}(M_{K^+K^-})\\
& +\frac{1}{3} G_{\eta\eta}(M_{K^+K^-})\tilde{t}_{\eta\eta \to K^+K^-}(M_{K^+K^-})\\
&+ \frac{1}{2} G_{\pi^0\pi^0} (M_{K^+K^-})\tilde{t}_{\pi^0\pi^0 \to K^+K^-}(M_{K^+K^-})  \\
& -\frac{2}{\sqrt{6}}G_{\pi^0\eta}(M_{K^+K^-})t_{\pi^0\eta \to K^+K^-}(M_{K^+K^-})\\
&+ G_{K^0\bar{K}^0}(M_{K^+K^-})t_{K^0\bar{K}^0\to K^+K^-} (M_{K^+K^-}) ]
\end{aligned}\label{iden}
\end{equation}
where we assume that $V_p$ is the same for \FIG{fig2a}(a) and (b), within the $SU(3)$ flavour symmetry, as in Refs.~\cite{Feng:2020jvp,Wang:2020pem}.
It should be noted that there is a factor of 2 in the terms associated with the identical particles $\pi^0\pi^0$ and $\eta\eta$ in Eq.~\ref{iden}. This has been canceled by the factor of $1/2$ in their propagators. Further details can be found in Ref.~\cite{Liang:2015qva}.
 The scattering matrix
$t_{i\to j}$ has been calculated as Eq.~\ref{TT}, and we take  $\tilde{t}_{\eta\eta\to i}=\sqrt{2}t_{\eta\eta\to i}$, $\tilde{t}_{\pi^0\pi^0 \to j}=\sqrt{2} t_{\pi^0\pi^0 \to j}$ for the two identical particles~\cite{Dias:2016gou,Wang:2020pem}.

The  amplitude of $\mathcal{T}_{2a}$ only contains the contribution from
isospin $I = 0$, and the $\mathcal{T}_{2b}$  can be decomposition with isospins $I=0$ and $I=1$ in the channels $K^+K^-$ and $K^0\bar{K}^0$ in $S$-wave,
\begin{equation}
\mathcal{T}_{2b} = \mathcal{T}^{I=0}_{2b}+\mathcal{T}^{I=1}_{2b},
\end{equation}
and it is easily done taking $G_{K^0\bar{K}^0}=G_{K^+K^-}$ as done in Refs.~\cite{Liang:2015qva,Wang:2020pem},
\begin{equation}
\begin{aligned}
\mathcal{T}^{I=0}_{2b} =& V_P V_{cd}V_{ud} \left[ G_{\pi^+\pi^-}(M_{K^+K^-}) t_{\pi^+\pi^-\to K^+K^-}(M_{K^+K^-}) \right.\\
& +\frac{1}{3}  G_{\eta\eta}(M_{K^+K^-})\tilde{t}_{\eta\eta \to K^+K^-}(M_{K^+K^-})\\
&+ \frac{1}{2} G_{\pi^0\pi^0}(M_{K^+K^-}) \tilde{t}_{\pi^0\pi^0 \to K^+K^-}(M_{K^+K^-})  \\
& +G_{K^0\bar{K}^0}(M_{K^+K^-})\left(\frac{1}{2}t_{K^0\bar{K}^0 \to K^+K^-}(M_{K^+K^-})\right.\\
&\left.\left. + \frac{1}{2} t_{K^+K^- \to K^+K^-}(M_{K^+K^-}) \right) \right]
\end{aligned}
\end{equation}
\begin{equation}
\begin{aligned}
\mathcal{T}^{I=1}_{2b} =& V_PV_{cd}V_{ud} \left[  -\frac{2}{\sqrt{6}}G_{\pi^0\eta}(M_{K^+K^-})t_{\pi^0\eta \to K^+K^-}(M_{K^+K^-})\right.\\
&+G_{K^0\bar{K}^0}(M_{K^+K^-})\left(\frac{1}{2}t_{K^0\bar{K}^0 \to K^+K^-}(M_{K^+K^-})\right.\\
&\left.\left.- \frac{1}{2} t_{K^+K^- \to K^+K^-} \right) \right].
\end{aligned}
\end{equation}

The $G_{MM}$ is the  loop function for the two mesons propagator ,
\begin{equation}
G_{MM}(s) = i\int \frac{d^4q}{(2\pi)^4}\frac{1}{(P-q)^2-m_1^2+i\epsilon}\frac{1}{q^2-m_2^2+i\epsilon},
\label{eq9}
\end{equation}
where $m_1$ and $m_2$ denote the masses of the two mesons, respectively, and $P$ and $q$ are the four-momentum of this meson-meson system and the second  meson in the center of mass system, respectively. The integral on $q$ in \Eq{eq9} is performed with a cutoff $|q_{\text{max}}| = 600$ MeV, as used in Ref.~\cite{Xie:2014tma,Liang:2014tia,Dias:2016gou}. 

For the \( I = 0 \) sector, there are five coupled channels: \(\pi^+ \pi^-(1)\), \(\pi^0 \pi^0(2)\), \(K^+ K^-(3)\), \(K^0 \bar{K}^0(4)\), and \(\eta \eta(5)\). The elements of the \(5 \times 5\) symmetric \(V\) matrix can be written as~\cite{Dias:2016gou} ,
\begin{equation}
\begin{aligned}
&V_{11} = -\frac{1}{2f^2} s, \qquad
V_{12} = -\frac{1}{\sqrt{2}f^2} (s - m_\pi^2), \\
&V_{13} = -\frac{1}{4f^2} s, \qquad
V_{14} = -\frac{1}{4f^2} s, \\
&V_{15} = -\frac{1}{3\sqrt{2}f^2} m_\pi^2, \qquad
V_{22} = -\frac{1}{2f^2} m_\pi^2, \\
&V_{23} = -\frac{1}{4\sqrt{2}f^2} s, \qquad
V_{24} = -\frac{1}{4\sqrt{2}f^2} s, \\
&V_{25} = -\frac{1}{6f^2} m_\pi^2, \qquad
V_{33} = -\frac{1}{2f^2} s, \qquad
V_{34} = -\frac{1}{4f^2} s, \\
&V_{35} = -\frac{1}{12\sqrt{2}f^2} \left(9s - 6m_\eta^2 - 2m_\pi^2\right), \\
&V_{44} = -\frac{1}{2f^2} s, \qquad
V_{45} = -\frac{1}{12\sqrt{2}f^2} \left(9s - 6m_\eta^2 - 2m_\pi^2\right), \\
&V_{55} = -\frac{1}{18f^2} \left(16m_K^2 - 7m_\pi^2\right).
\end{aligned}
\end{equation}

Besides, for the $I = 1$ sector, three channels are coupled, $K^+K^-(1), K^0\bar{K}^0(2), \pi^0\eta(3)$, and the
elements of the $V$ can be obtained as follows~\cite{Xie:2014tma}
\begin{equation}
\begin{aligned}
&V_{K^{+}K^{-}\to K^{+}K^{-}} = -\frac{1}{2f^2}s, \qquad V_{K^{+}K^{-}\to K^{0}\bar{K}^{0}} = -\frac{1}{4f^2}s,\\
&V_{K^{+}K^{-}\to\pi^{0}\eta} = \frac{- \sqrt{3}}{12f^2} \left( 3s - \frac{8}{3}m_K^2 - \frac{1}{3}m_{\pi}^2 - m_{\eta}^2 \right),\\
&V_{K^{0}\bar{K}^{0}\longrightarrow K^{0}\bar{K}^{0}} = -\frac{1}{2f^2}s, \\
&V_{K^{0}\bar{K}^{0}\longrightarrow\pi^{0}\eta}
= -V_{K^{+}K^{-}\to\pi^{0}\eta}, \qquad V_{\pi^{0}\eta\to\pi^{0}\eta} = -\frac{m_{\pi}^2}{3f^2}\\
\end{aligned}
\end{equation}
where $f=f_{\pi}$ is the pion decay constant, $m_{\pi}$ and $m_K$ are
the averaged masses of the pion and kaon, respectively. 

Furthermore, the effective energy range in the ChUA
has limitations. In order to make reliable predictions up to higher energy region, we need
to expand the scattering amplitudes above the energy cut $\sqrt{s_{\text{cut}}}=1050$ MeV smoothly~\cite{Debastiani:2016ayp}
\begin{equation}
G(s)T(s) = G(s_{\text{cut}})T(s_{\text{cut}})e^{-\alpha\left(\sqrt{s}-\sqrt{s_{\text{cut}}}\right)}, \quad \sqrt{s} > \sqrt{s_{\text{cut}}}
\label{alpha}
\end{equation}
where we take the value $\alpha = 0.0037$  $\text{MeV}^{-1}$.

\subsection{Invariant Mass Distributions}

Using the theoretical formalism shown above, the total decay amplitude for the process of $\Xi_c^+\to \Sigma^+K^+K^-$ can be written as:
\begin{equation}
	|\mathcal{T}|^2 = |\mathcal{T}^{\Xi(1/2^-)}+\mathcal{T}^{I=0}+\mathcal{T}^{I=1}|^2.
	\label{eq19}
\end{equation}
Here, $\mathcal{T}^{I=0} = \mathcal{T}_{2a} + \mathcal{T}_{2b}^{I=0}$, while $\mathcal{T}^{I=1}$ denotes $\mathcal{T}_{2b}^{I=1}$.

The double differential width can be calculated using the following equation~\cite{ParticleDataGroup:2022pth}:
\begin{eqnarray}
\frac{d^2\Gamma}{dM^2_{K^+K^-}{dM^2_{K^-\Sigma^+}}}&=&\frac{4M_{\Xi_c^+}M_{\Sigma^+}}{{(2\pi)}^3{32M_{\Xi_c^+}}^3}|\mathcal{T}|^2. \label {eq:dgammadm12dm23}  \nonumber \\
\end{eqnarray}
For a given value of $M_{12}$, the range of $M_{23}$ is constrained by the following condition:
\begin{eqnarray}
	M^{\rm max}_{23} &= &\sqrt{\left(E_{2}^\ast+E_{3}^\ast\right)^2 -\left(\sqrt{E_{2}^{\ast2}-m_{2}^2}-\sqrt{E_{3}^{\ast2}-m_{3}^2}\right)^2}, \nonumber \\
	M_{23}^{\rm min} &=&\sqrt{\left(E_{2}^\ast+E_{3}^\ast\right)^2 -\left(\sqrt{E_{2}^{\ast2}-m_{2}^2}+\sqrt{E_{3}^{\ast2}-m_{3}^2}\right)^2},\nonumber \\
\end{eqnarray}
where $E_{2}^\ast$ and $E_{3}^\ast$ are the energies of particles 2 and 3 in the $M_{12}$ rest frame, which can be written as
\begin{align}
	&E_{2}^\ast=\frac{M_{12}^2-m_{1}^2+m_{2}^2}{2M_{12}},  \nonumber \\
	&E_{3}^\ast=\frac{M_{\Xi_c^+}^2-M_{12}^2-m_{3}^2}{2M_{12}},
\end{align}
with $m_1$, $m_2$ and $m_3$ are the masses of involved particles $1, 2$, and 3, respectively.

\section{results and discussions}
\label{III}
\begin{figure}[tbhp]
	\centering
	\includegraphics[scale=0.45]{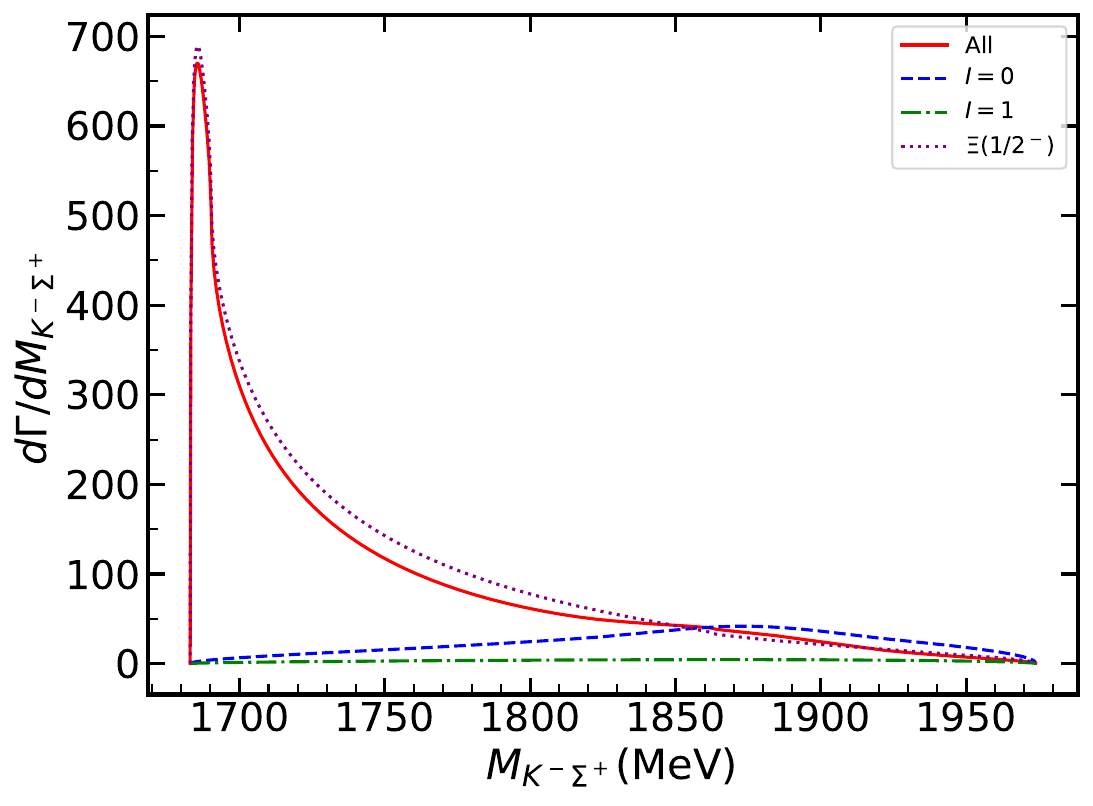}
	\caption{The $K^-\Sigma^+$ invariant mass distribution of the process $\Xi_c^+ \to \Sigma^+K^+K^-$ decay.}
	\label{m23}	
\end{figure}
\begin{figure}[tbhp]
	\centering
	\includegraphics[scale=0.45]{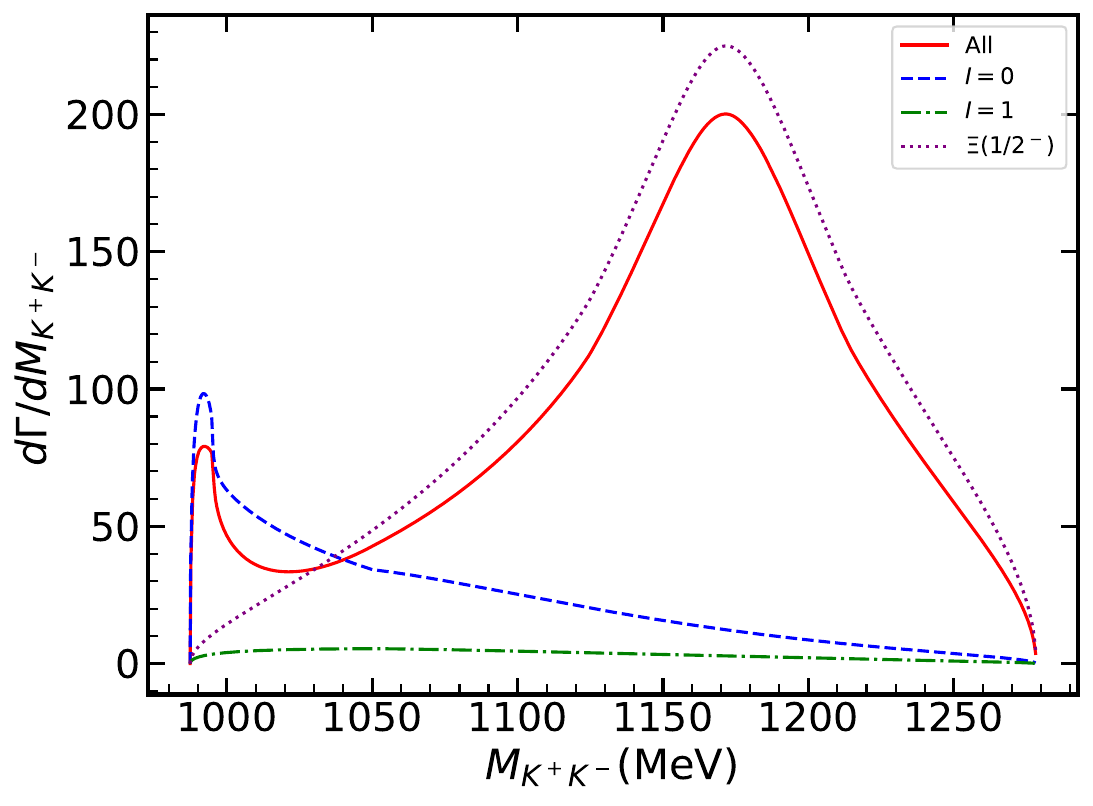}
	\caption{The $K^+K^-$ invariant mass distribution of the process $\Xi_c^+ \to \Sigma^+K^+K^-$ decay.}
	\label{m13}	
\end{figure}

This section presents the results obtained based on the theoretical formalism introduced in Sec.~\ref{II}. 
\begin{figure}[tbhp]
	\centering
	\includegraphics[scale=0.45]{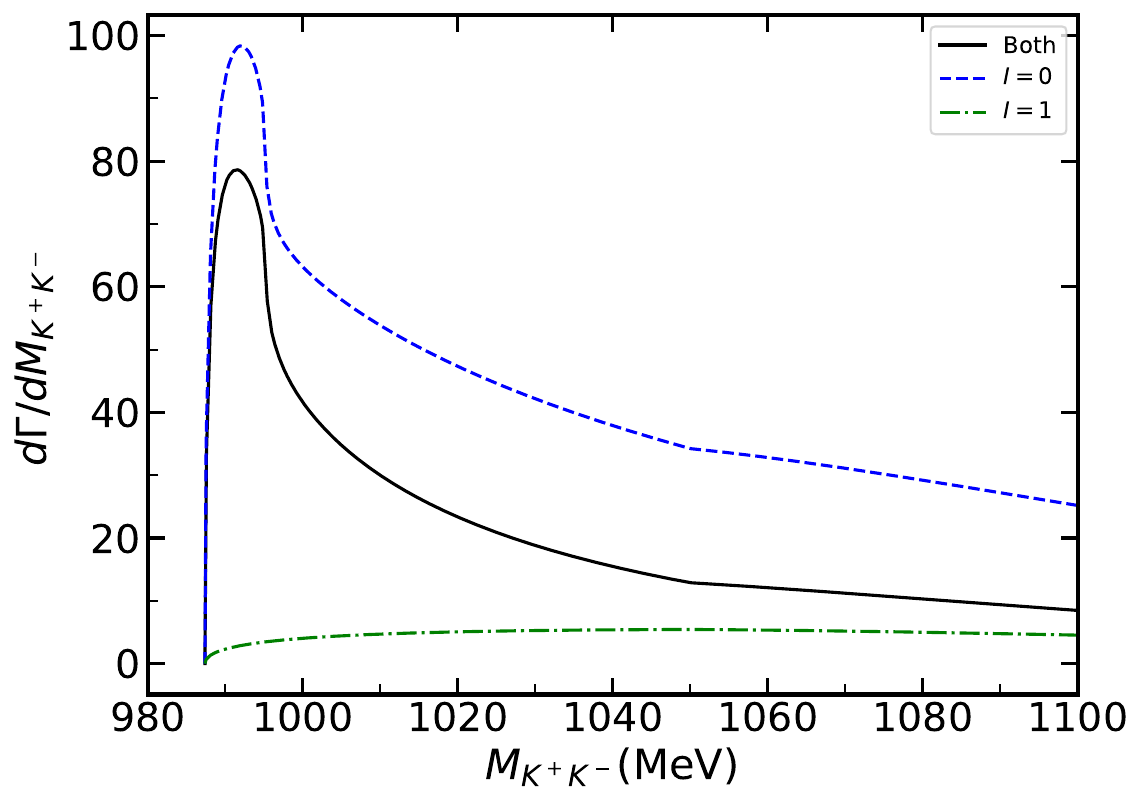}
	\caption{The $K^+K^-$ invariant mass distributions for the $\Xi_c^+\to \Sigma^+K^+K^-$ decay.  The curves labeled as ‘$I = 0$’, ‘$I = 1$’ and ‘Both’ correspond to the contributions from the $\mathcal{T}^{I=0}$, $\mathcal{T}^{I=1}$, and  $(\mathcal{T}^{I=0}+\mathcal{T}^{I=1})$ respectively. }
	\label{minI}	
\end{figure}
\begin{figure}[tbhp]
	\centering
	\includegraphics[scale=0.45]{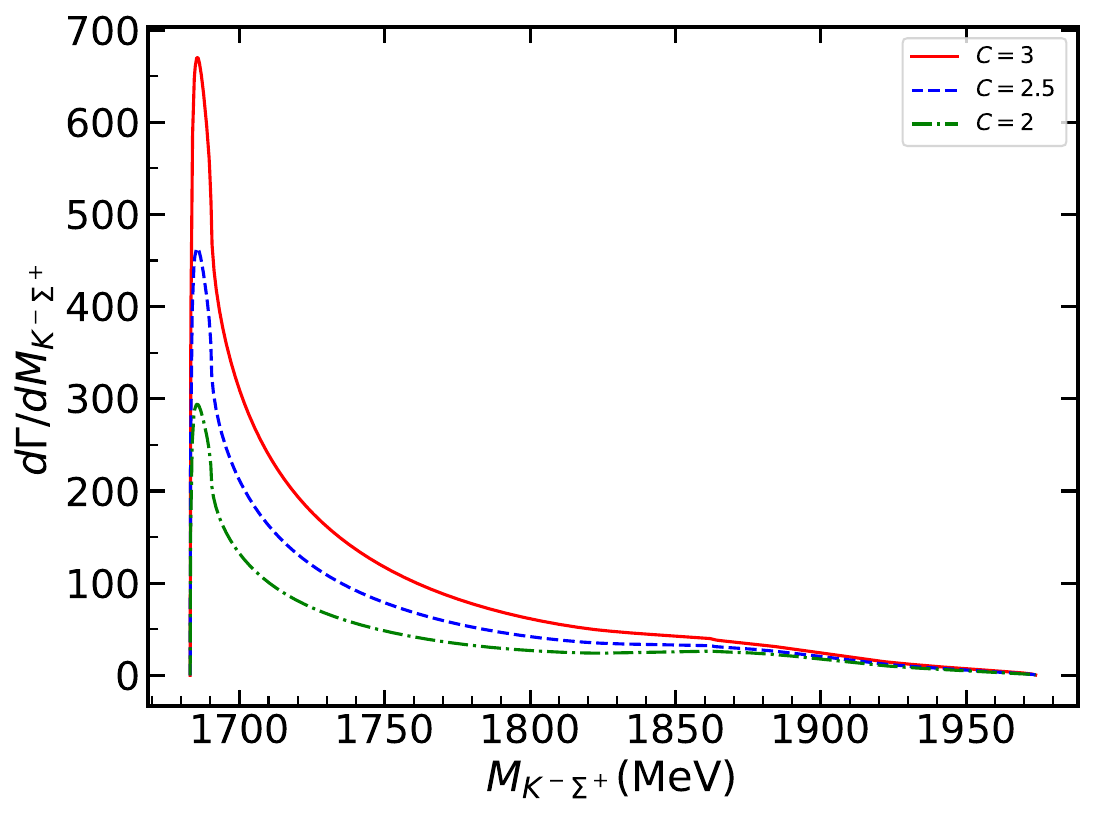}
	\includegraphics[scale=0.45]{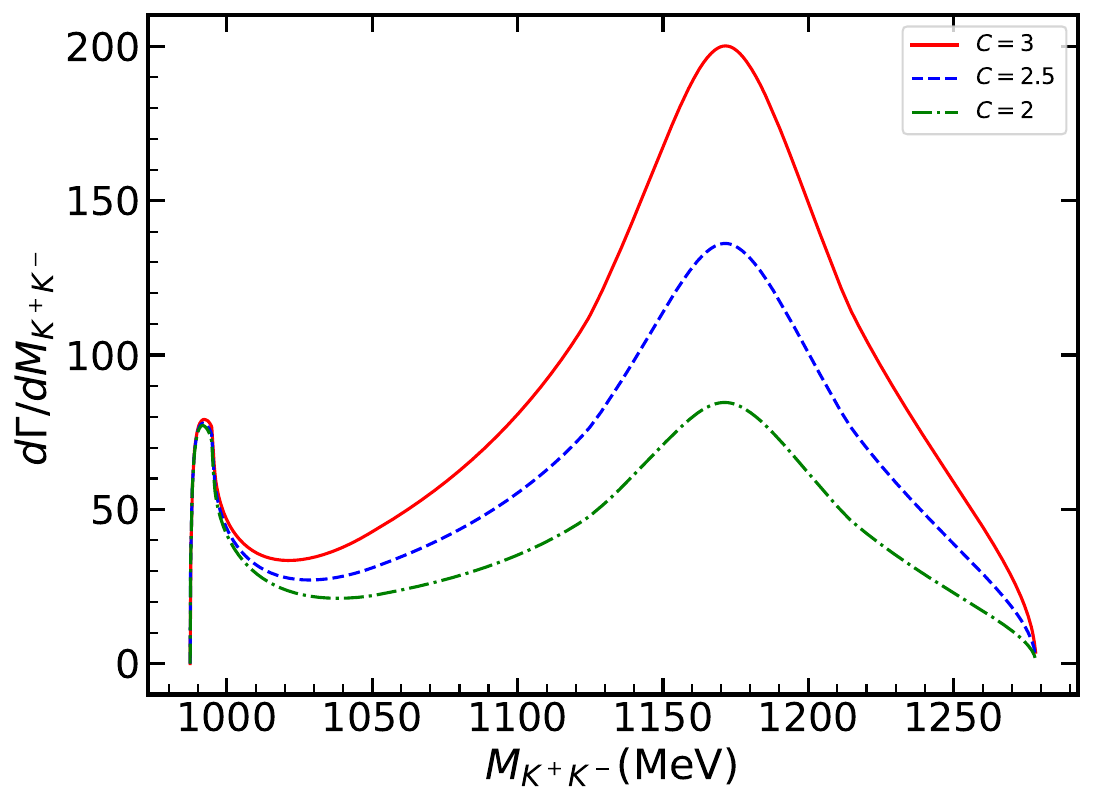}
	\caption{Distributions of $K^-\Sigma^+$ and $K^+K^-$ invariant masses of $\Xi_c^+ \to \Sigma^+K^+K^-$ decay with the different values of color factor $C$.}
	\label{facc}	
\end{figure}
The model incorporates two free parameters: $C$ and $V_p$. The color factor $C$ quantifies the relative contribution of external $W^+$ emission with respect to internal $W^+$ emission. Its initial value is set to $C=3$, and its impact on the overall results will be analyzed subsequently. A global normalization factor $V_p=1$ is employed in presenting the results under an arbitrary normalization scheme, since this parameter does not affect the shape of the invariant mass distribution of the final state. 

\begin{figure}[tbhp]
	\centering
	\includegraphics[scale=0.45]{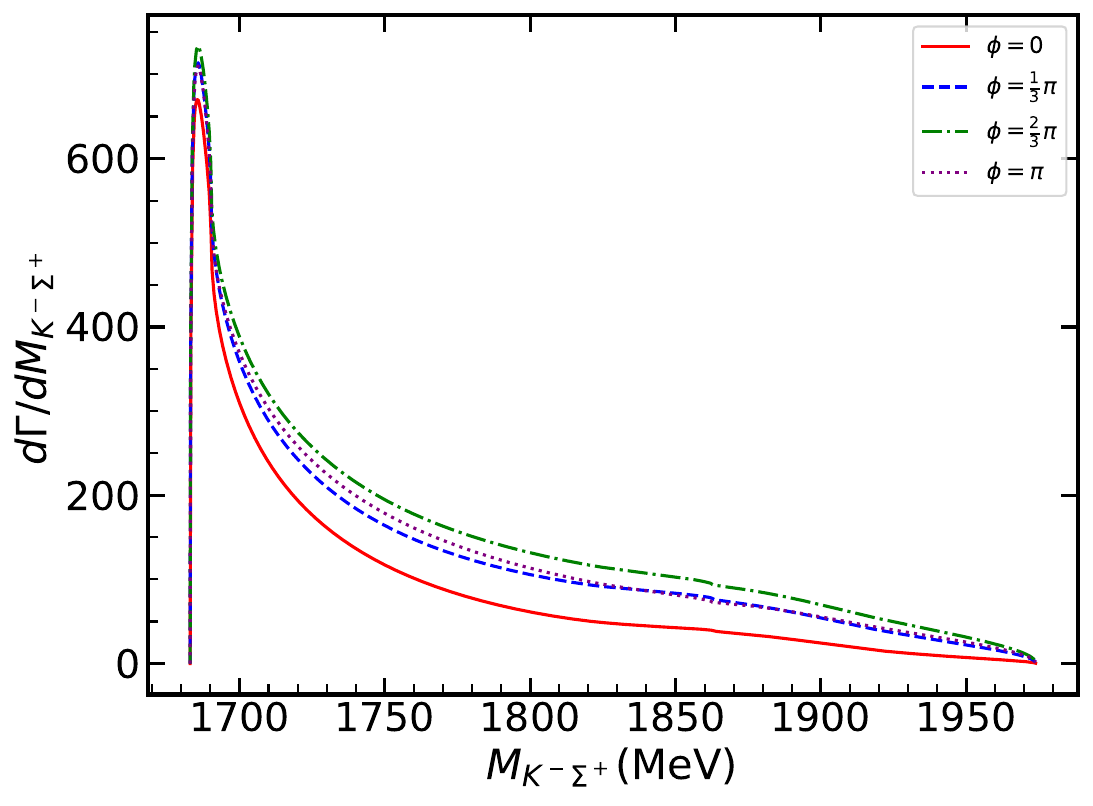}
	\caption{The $K^-\Sigma^+$ invariant mass distributions of
		the process  $\Xi_c^+ \to \Sigma^+K^+K^-$ decay with the interference phase $\phi = 0, \frac{1}{3}\pi, \frac{2}{3}\pi$ and $\pi$, respectively.}
	\label{phi23}	
\end{figure}
\begin{figure}[tbhp]
	\centering
	\includegraphics[scale=0.45]{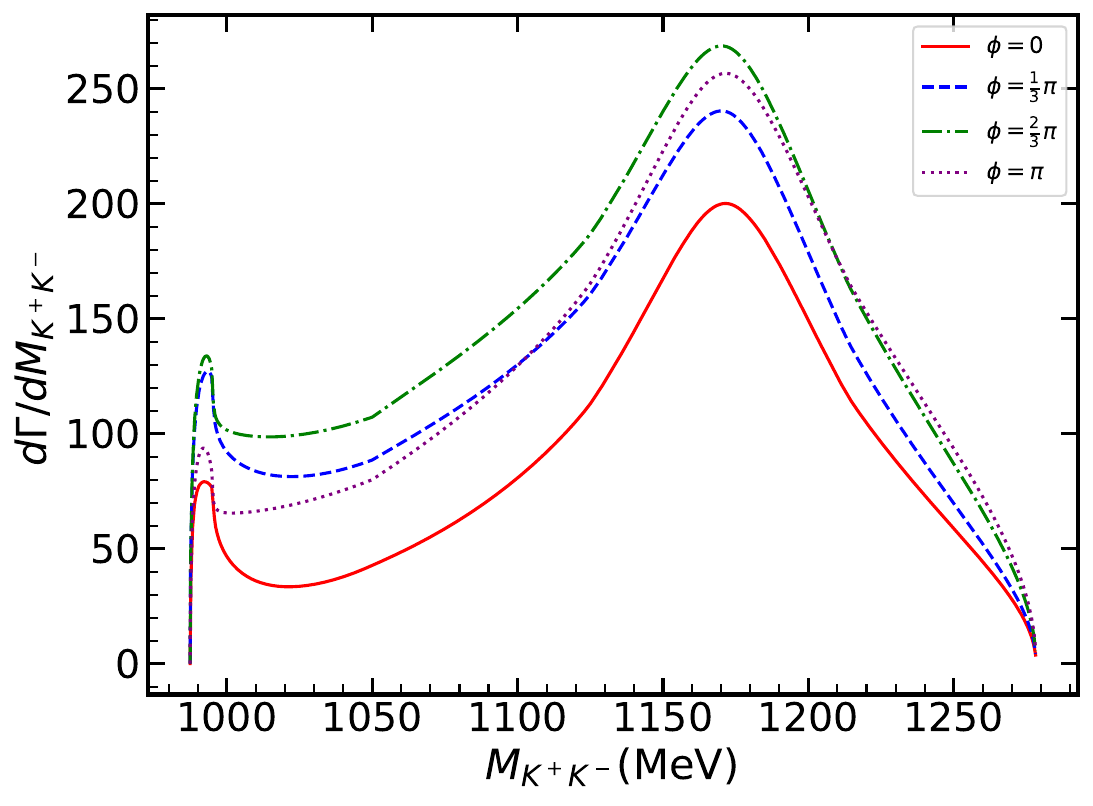}
	\caption{The $K^+K^-$ invariant mass distributions of
		the process  $\Xi_c^+ \to \Sigma^+K^+K^-$ decay with the interference phase $\phi = 0, \frac{1}{3}\pi, \frac{2}{3}\pi$ and $\pi$, respectively.}
	\label{phi13}	
\end{figure}

\begin{figure}[tbhp]
	\centering
	\includegraphics[scale=0.45]{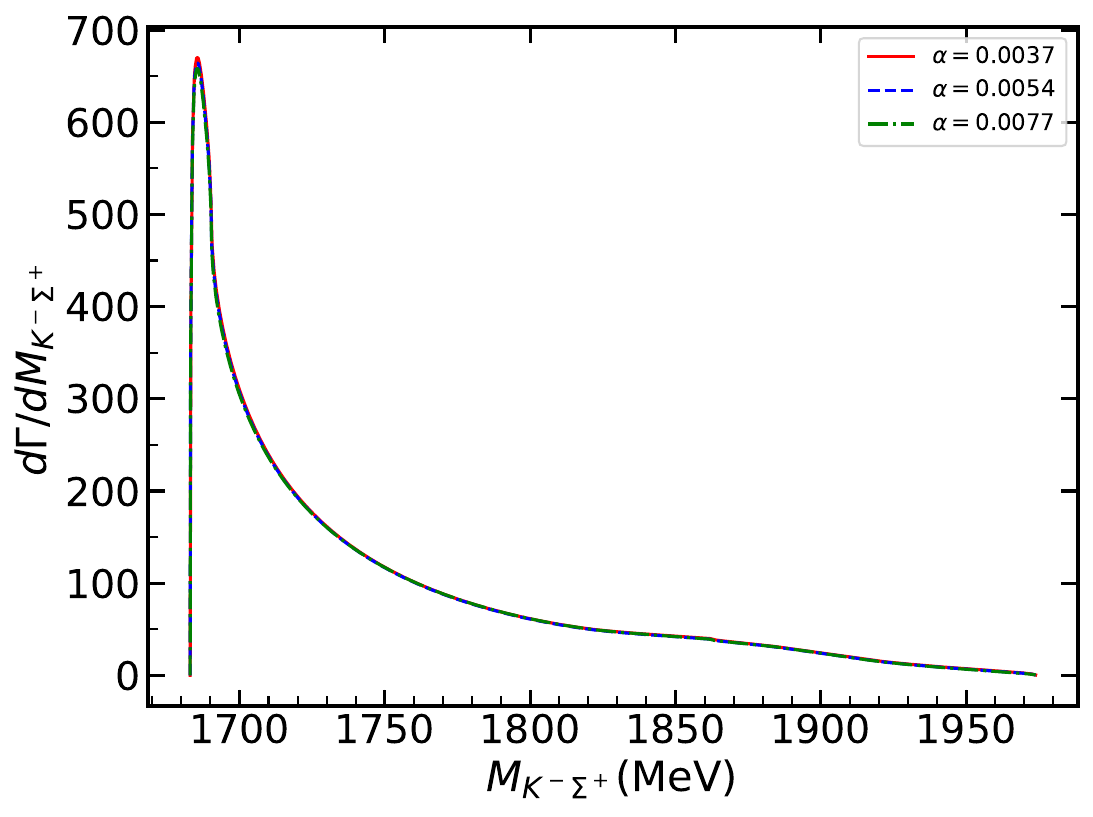}
	\includegraphics[scale=0.45]{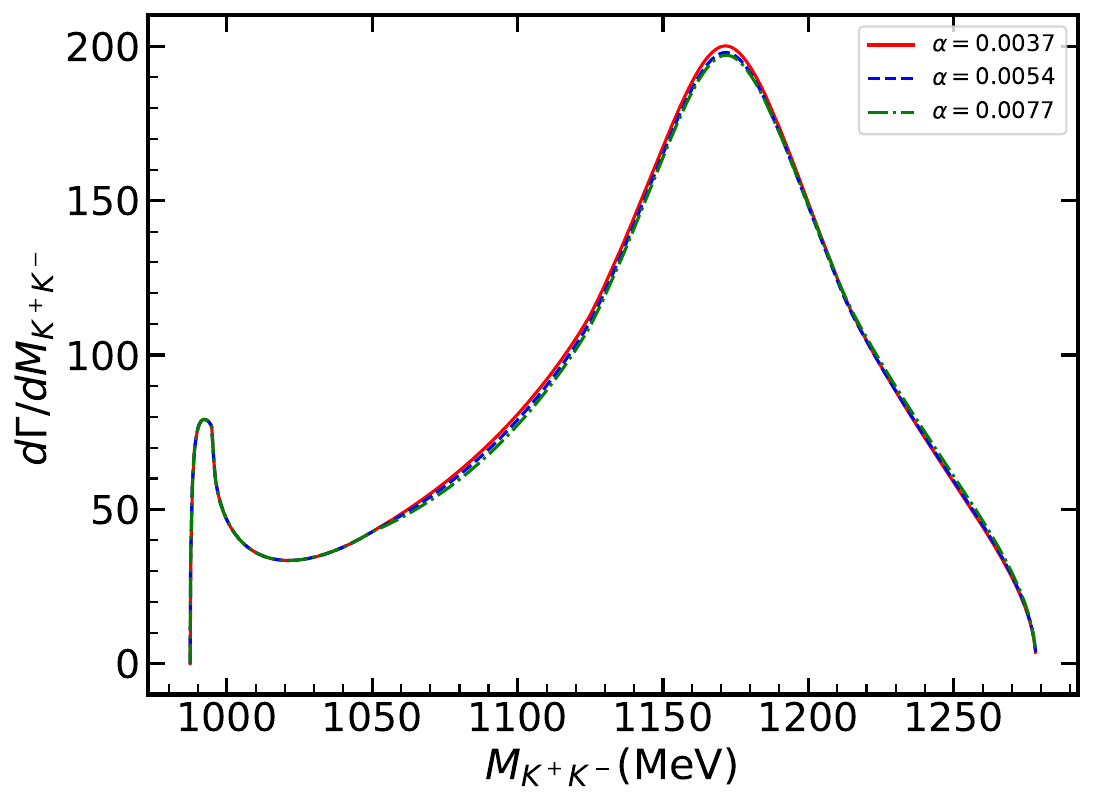}
	\caption{Distributions of $K^-\Sigma^+$ and $K^+K^-$ distribu-\\tions with the different values of $\alpha$ in the smooth factor.}
	\label{alp}	
\end{figure}

In \FIG{m23} and \FIG{m13}, we show the $K^-\Sigma^+$ and $K^+K^-$ the mass distributions of the process $\Xi_c^+ \to \Sigma^+K^+K^-$. The red line represents the contribution to the total amplitude,
while the blue dashed line and green dot dashed line denote the contributions from the
the contributions from the isospin $I=0$ and $I=1$  amplitudes, respectively. The purple dotted line reflects the contribution from the intermediate state $\Xi(1/2^-)$.
A distinct peak is observed in the $K^-\Sigma^+$ invariant mass distribution, associated with the $\Xi(1/2^-)$ resonance state. One can find the peak position near 1690 MeV and the narrow width about 30 MeV in the $K^-\Sigma^+$ invariant mass distribution are compatible with the properties of the $\Xi(1690)$ resonance in the PDG~\cite{ParticleDataGroup:2024cfk}.
 Furthermore, a cusp structure at approximately 980 MeV in the $K^+K^-$ invariant mass distribution corresponds to the effects
of the isospin $I = 0$ and $I = 1$ contributions in the rescattering contributions, which should be the reflection of the $f_0(980)$ and $a_0(980)$ resonance states. 
A detailed comparison of the isospin $I=0$ and $I=1$ contributions is presented in \FIG{minI}, focusing on the region around 980 MeV in the $K^+K^-$ invariant mass distribution. From the figure, we can see that, at low energies, the contribution from the isospin $I=1$ is much smaller than that from the isospin $I=0$, and the interference between them is destructive.

Subsequently, we explored the impact of varying color factor $C$ values ($C=2$, $2.5$, and 3) on the results. As shown in the invariant mass distributions of $K^-\Sigma^+$ and $K^+K^-$ in Fig.~\ref{facc}, we can clearly see that the $C$ factor value only affected the the strength of signals, but the signals remain clearly visible.

Additionally, due to possible phase interference among different terms in Eq.~\ref{eq19}, we introduce a phase factor $e^{i\phi}$ to the amplitude $\mathcal{T}^{\Xi(1/2^-)}$ in Eq.~\ref{eq19} and examine its influence on the final results. We perform calculations for different values of $\phi = 0, \frac{1}{3}\pi, \frac{2}{3}\pi$ and $\pi$, with the results displayed in Figs. \FIG{phi23} and \FIG{phi13}.

It can be seen that although interference from different phase angles $\phi$ could distort the line shapes, the near threshold enhancement structure associated with the $\Xi(1/2^-)$ state can still be clearly observable in the $K^-\Sigma^+$ invariant mass distribution, and the peak structure near $980$ MeV related to the $f_0(980)$ and $a_0(980)$ resonances remains evident in the $K^+K^-$ distribution.

Finally, we tested the dependence of our results on
	the parameter $\alpha$ in smooth factor of \Eq{alpha}. 
	In Ref.~\cite{Debastiani:2016ayp}, the three values $\alpha = 0.0037$, $0.0054$, and $0.0077$ MeV$^{-1}$, are chosen to achieve approximately factors of 3, 5, and 10 suppression of $Gt$ at $\sqrt{s} =\sqrt{s_{\text{cut}}} + 300$ MeV, respectively. So we adopt these same values of $\alpha$ to calculate the invariant mass distributions, with the results presented in \FIG{alp}. From
	the invariant mass distributions of $M_{K^-\Sigma^+}$ and $M_{K^+K^-}$ in
	\FIG{alp}, the position of the peaks is not affected by the value of $\alpha$, which has only a minor effect on the results.

 \section{SUMMARY}
 \label{IV}

Studying charmed baryon non-leptonic weak decays is an useful method for investigating low-lying excited baryons and exploring the properties of light baryons with quantum numbers $J^P = 1/2^-$.

In this work, we investigate the process $\Xi_c^+ \to \Sigma^+K^+K^-$ within the chiral unitary approach. The intermediate $\Xi(1/2^-)$ resonance is dynamically generated via the $S$-wave interaction between pseudoscalar mesons and baryons, taking into account the coupled channels $K^-\Sigma^+$, $\bar{K}^0\Sigma^0$, $\bar{K}^0\Lambda$, $\pi^+\Xi^-$, $\pi^0\Xi^0$, and $\eta\Xi^0$. Meanwhile, the $f_0(980)$ and $a_0(980)$ states are dynamically
generated through the $S$-wave pseudoscalar meson–pseudoscalar meson interactions.
According to our calculations, a clear peak can be seen at around 1690 MeV in the invariant mass distributions of $K^-\Sigma^+$, which could be related to the $\Xi(1/2^-)$.
Also, we have demonstrated the dependence of the relative phase angle between amplitudes, and discussed the influence of the different values of the color factor $C$. 
This indicates that the $\Xi(1/2^-)$ signal remains clearly distinct in the $K^-\Sigma^+$ invariant mass distribution, and the role of $f_0(980)$ and $a_0(980)$ exhibits significant effects in the low $K^+K^-$ invariant mass region.

We hope that further experimental measurements will verify the properties of the $\Xi(1/2^-)$, $f_0(980)$, and $a_0(980)$ resonances predicted by our model.

 \begin{acknowledgments}
 	\noindent
HS is supported by the National Natural Science Foundation of China (Grant No.12075043). XL is supported
by the National Natural Science Foundation of China under Grant No.12205002.
 \end{acknowledgments}

\bibliography{ref}
\end{document}